\documentclass[a4paper,11pt]{article}
\pdfoutput=1 

\usepackage{jcappub} 

\usepackage[T1]{fontenc} 
\usepackage{wasysym} 
\usepackage{multirow}

\title{\boldmath Beta/gamma and alpha backgrounds in CRESST-II Phase 2}

\author[a,1]{R. Strauss,\note{Corresponding author.}}

\author[a]{G. Angloher}

\author[c]{A. Bento}

\author[d]{C. Bucci}

\author[d]{L. Canonica}

\author[b,e]{A. Erb}

\author[b]{F.v. Feilitzsch}

\author[a]{N. Ferreiro\,\,Iachellini}

\author[d]{P. Gorla}

\author[f]{A. G\"utlein}

\author[a]{D. Hauff}

\author[g]{J. Jochum}

\author[a]{M. Kiefer}

\author[f]{H. Kluck}

\author[h]{H. Kraus}

\author[b]{J.-C. Lanfranchi}

\author[g]{J. Loebell}

\author[b]{A. M\"unster}

\author[a]{F. Petricca}

\author[b]{W. Potzel}

\author[a]{F. Pr\"obst}

\author[a]{F. Reindl}

\author[b]{S. Roth}

\author[g]{K. Rottler}

\author[g]{C. Sailer}

\author[d]{K. Sch\"affner}

\author[f]{J. Schieck}

\author[g]{S. Scholl}

\author[b]{S. Sch\"onert}

\author[a]{W. Seidel}

\author[b,2]{M.v. Sivers\note{Present address: Albert Einstein Center for Fundamental Physics, University of Bern, CH-3012 Bern, Switzerland.}}

\author[a]{L. Stodolsky}

\author[g]{C. Strandhagen}

\author[a]{A. Tanzke}

\author[g]{M. Uffinger}

\author[b]{A. Ulrich}

\author[g]{I. Usherov}

\author[b]{S. Wawoczny}

\author[b]{M. Willers}

\author[a]{M. W\"ustrich}

\author[b]{A. Z\"oller}

\affiliation[a]{Max-Planck-Institut f\"ur Physik,   D-80805 M\"unchen, Germany}
\affiliation[b]{Physik-Department, Technische Universit\"at M\"unchen, D-85748 Garching, Germany}
\affiliation[c]{CIUC, Departamento de Fisica, Universidade de Coimbra, P3004 516 Coimbra, Portugal}
\affiliation[d]{INFN, Laboratori Nazionali del Gran Sasso, I-67010 Assergi, Italy}
\affiliation[e]{Walther-Mei\ss ner-Institut f\"ur Tieftemperaturforschung,  D-85748 Garching, Germany}
\affiliation[f]{Institut f\"ur Hochenergiephysik der \"Osterreichischen Akademie der Wissenschaften, A-1050 Wien, Austria and Atominstitut, Vienna University of Technology, A-1020 Wien, Austria}
\affiliation[g]{Physikalisches Institut, Eberhard-Karls-Universit\"at T\"ubingen,   D-72076 T\"ubingen, Germany}
\affiliation[h]{Department of Physics, University of Oxford, Oxford OX1 3RH, United Kingdom}

\emailAdd{strauss@mpp.mpg.de}

\abstract{The experiment CRESST-II aims at the detection of dark matter with scintillating CaWO$_4$ crystals operated as cryogenic detectors. Recent results on spin-independent WIMP-nucleon scattering from the CRESST-II Phase 2 allowed to probe a new region of parameter space for WIMP masses below 3\,GeV/c$^2$. This sensitivity was achieved after   background levels were reduced significantly.  We present extensive background studies of a CaWO$_4$ crystal, called TUM40,  grown at the Technische Universit\"at M\"unchen. The average beta/gamma rate of  3.51/[kg\,keV\,day] (1-40\,keV) and the total intrinsic alpha activity from natural decay chains of   $3.08\pm0.04$\,mBq/kg are the lowest  reported for CaWO$_4$ detectors. Contributions from  cosmogenic activation, surface-alpha decays, external radiation and intrinsic alpha/beta emitters are investigated in detail. A Monte-Carlo based background decomposition allows to identify the origin of the majority of  beta/gamma events in the energy region relevant for dark matter search.  }

\begin{document}
\maketitle
\flushbottom

\section{Introduction}
During the last two decades, the sensitivity of experiments aiming at the direct detection of particle dark matter \cite{Bertone:2004pz} in form of weakly interacting massive particles (WIMPs) \cite{Jungman:1995df}  has been constantly improved. For the spin-independent WIMP-nucleon cross section impressive sensitivities were reached: currently the liquid-xenon based LUX \cite{Akerib:2013tjd} experiment reports the best upper limit ($7.6\cdot 10^{-10}$\,pb at 33\,GeV/c$^2$). A variety of experiments with different techniques \cite{Cushman:2013zza} have been operated, however, the results are not  consistent. A few experiments \cite{Bernabei:2010mq,Aalseth:2012if,PhysRevLett.111.251301}, among which is  CRESST-II (Cryogenic Rare Event Search with Superconducting Thermometers)  \cite{Angloher:2014myn}, reported a signal excess which is not compatible with limits of other dark matter searches \cite{Aprile:2012nq,Agnese:2014aze,Agnese:2013jaa,Armengaud:2012pfa}. Data from a re-analysis of the commissioning run of CRESST-II \cite{PhysRevD.85.021301} showed slight tension with a WIMP interpretation of CRESST-II data and, recently, the first data of  CRESST-II Phase 2 \cite{Angloher:2014myn} suggest a background origin of the excess.\\
The improvement in sensitivity of CRESST-II detectors has been achieved by a significant  reduction of backgrounds. In this paper, we present a comprehensive study of the different backgrounds observed in CRESST-II Phase 2 (section \ref{sec:experiment}). The knowledge of the background origin is crucial for  future dark matter searches based on the CRESST technology. The Monte-Carlo (MC) based decomposition of the spectrum into different background sources which is presented in section \ref{sec:decomposition} gives an important input for future R\&D activities.   
%

\section{The detector module TUM40}\label{sec:TUM40}

CRESST-II detectors are based on a two-channel detector readout which is the key feature to discriminate irreducible radioactive backgrounds. CaWO$_4$ crystals of 200-300\,g each, equipped with transition-edge-sensors \cite{Angloher2009270},  are operated as cryogenic detectors (called phonon detectors) which allow to measure precisely the total deposited energy $E$ of a particle interaction. An excellent energy threshold of $\mathcal{O}$(500\,eV) and a resolution on a \permil level (at 2.6\,keV) were achieved \cite{Angloher:2014myn,strauss:2014part1}. In addition, the scintillation light output of these crystals is monitored by a  cryogenic silicon-on-sapphire  detector (called light detector). Since the relative amount of scintillation light, called light yield ($LY$)\footnote{The light yield is defined as 1 for electron recoils of 122\,keV, i.e. the light energy of such events is defined as 122\,keV$_{\mathrm{ee}}$, the so-called electron-equivalent energy \cite{strauss:2014part1}.} , strongly depends  on the kind of particle interaction (due to quenching \cite{birks1964theory}) this  channel provides a discrimination of beta/gamma ($LY\sim1$), alpha ($LY\sim0.2$) and nuclear-recoil events ($LY\lesssim0.1$). To a certain extent even O, Ca and W recoils can be distinguished \cite{Strauss:2014ab}. CRESST uses a unique multi-element target for WIMP search in a single experiment.\\
Due to a finite resolution of the light channel, the beta/gamma and nuclear-recoil bands overlap in the region-of-interest (ROI) for dark matter search which is typically defined  between energy threshold and 40\,keV. Several background sources related to surface-alpha decays were identified in the previous run of CRESST-II \cite{Angloher:2012vn}. In particular recoiling $^{206}$Pb nuclei from $^{210}$Po decays were a serious background. These events appear at low light yields similar to W recoils \cite{strauss:2014part1} and might thus be indistinguishable from potential WIMP scatters.\\
However, the  decay of $^{210}$Po has a corresponding alpha particle with an energy in the MeV-range which is used to reject this  background source. Alphas hitting scintillating materials surrounding the CaWO$_4$ crystals produce sufficient additional light to clearly identify a simultaneously occurring recoil of a heavy nucleus.\\
\begin{figure}
\centering
\includegraphics[width=0.7\textwidth]{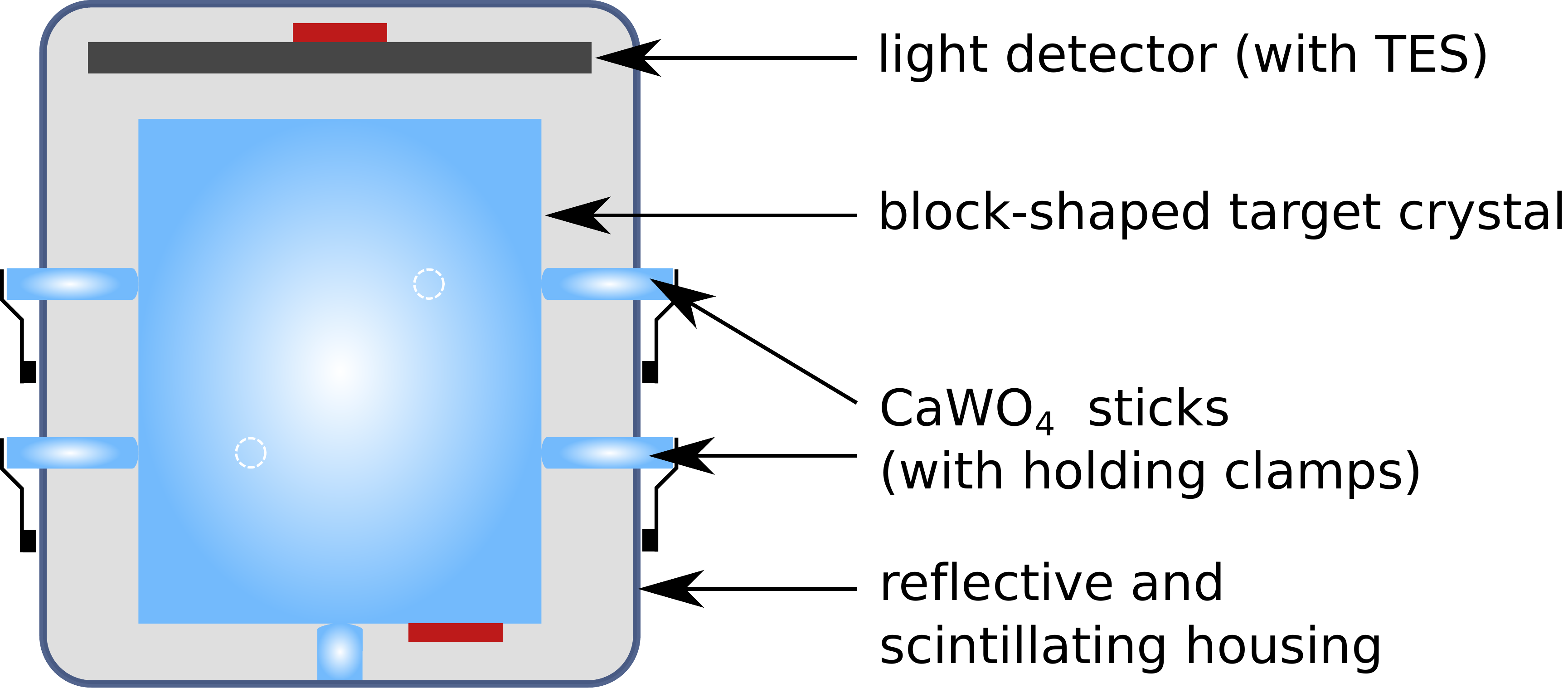}
\caption{Schematic view of the novel detector module. The block-shaped target crystal TUM40 with a mass of $m=249$\,g  is held by CaWO$_4$ sticks. Together with the scintillating polymeric foil the sticks establish a fully scintillating inner detector housing which provides an efficient active veto against surface events. A separate silicon-on-sapphire light detector is installed. For details see \cite{strauss:2014part1}.  }
\label{fig:stickHolder}       
\end{figure}
The crystal investigated in this paper, a block-shaped CaWO$_4$ crystal of 249\,g, is mounted in a novel detector module.  Instead of (non-scintillating) metal clamps, CaWO$_4$ sticks are used to support the target crystal. The crystal is surrounded by a reflective and scintillating polymeric foil which together with the CaWO$_4$ sticks realizes a detector housing with fully-scintillating inner surfaces. A schematic view of the novel detector design is shown in Fig. \ref{fig:stickHolder}, for a detailed  description  see \cite{strauss:2014part1}. As will be presented in section \ref{sec:surface}, this concept provides an efficient rejection of surface-alpha induced Pb-recoils and solves the main background problem of earlier CRESST-II runs \cite{Angloher:2012vn}. With this detector, a phonon trigger threshold of $\sim0.60$\,keV and a resolution of $\sigma{=}(0.090{\pm}0.010)\,$keV (at 2.60\,keV) were achieved \cite{strauss:2014part1,Angloher:2014myn}.\\
The detector module presented in this paper uses a CaWO$_4$ crystal, called TUM40,  grown at the crystal laboratory of the  Technische Universit\"at M\"unchen (TUM) \cite{erb}. At this institute, a Czochralski crystal-production facility was set up, dedicated to the growth of CaWO$_4$. The furnace is exclusively used for the CRESST experiment which is crucial for the requirements in terms of radiopurity.  Since radioactive isotopes in CaWO$_4$ crystals  can originate from the raw materials themselves \cite{muenster_2014}, a careful selection of these materials is necessary.  In particular, in CaCO$_3$ powders which are often extracted from geological settings considerable contaminations were found (e.g. $^{226}$Ra with an activity of $\mathcal{O}$(10\,mBq/kg) \cite{muenster_2014}).   Beyond that, impurities can be introduced during processing and handling of raw materials, during crystal growth and  detector production. An important source of contamination is Rn-implantation from air. The gaseous isotope $^{222}$Rn (half-life 3.6\,d) which is produced in the $^{238}$U chain deposits onto surfaces and the daughter isotopes can be implanted. Therefore, in particular the surfaces of the detectors and of their surroundings  get contaminated and the exposure to air containing Rn has to be minimized. Further, storing CaWO4, which is often delivered as a powder, requires adequate storage to minimize Rn-related backgrounds.  \\

\section{Experimental results}\label{sec:experiment}
The first data of TUM40 in CRESST-II Phase 2 with an exposure of 29\,kg-days  were used for a low-mass WIMP analysis \cite{Angloher:2014myn} and for the background studies presented in this paper.

\subsection{Beta/gamma background}\label{sec:low_energy_spectrum}
\begin{figure}
\centering
\includegraphics[width=0.9\textwidth]{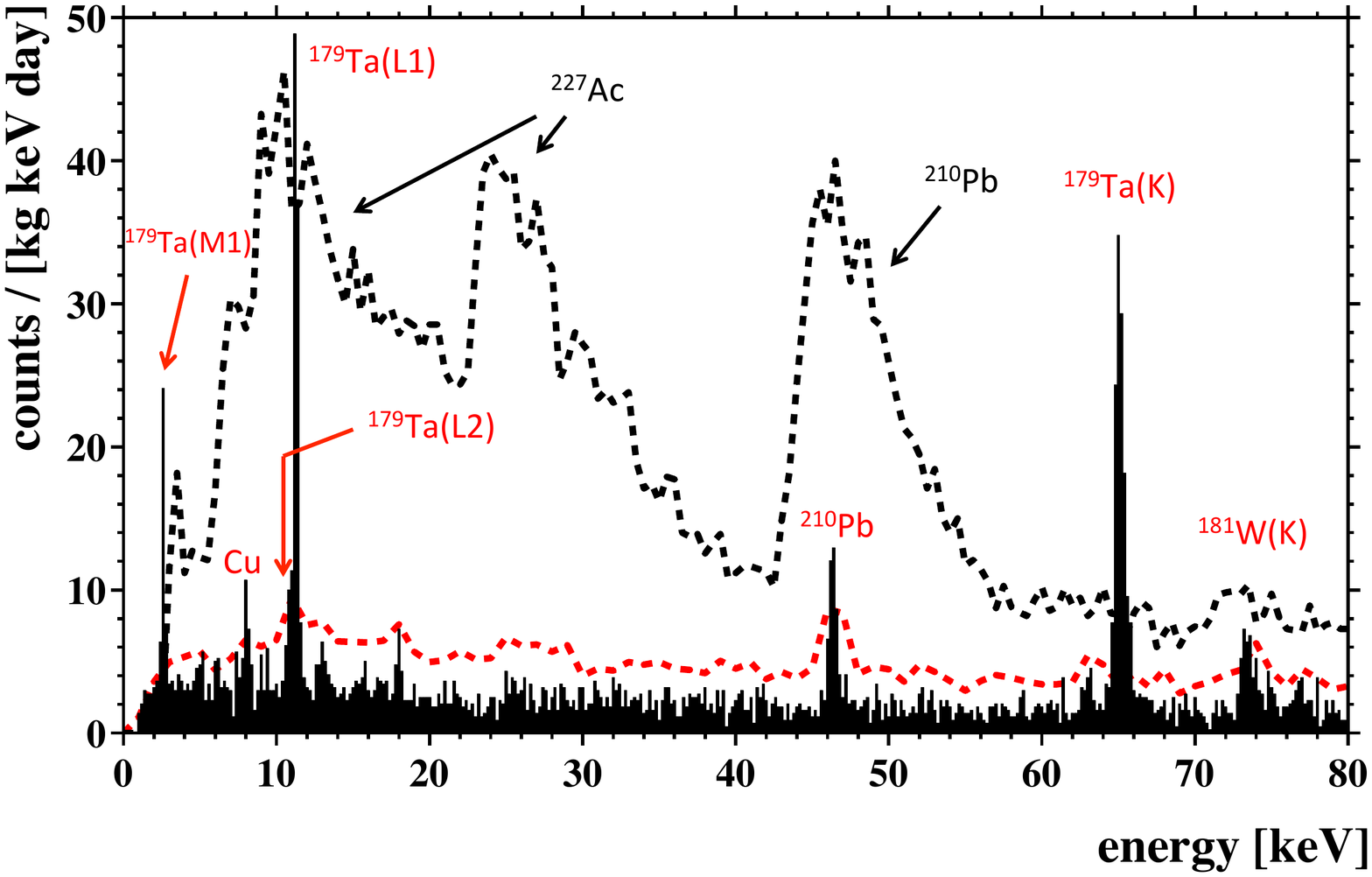}
\caption{Histogram of the low-energy events  (black bars) recorded during CRESST-II Phase 2 with TUM40 which was produced at the Technische Universit\"at M\"unchen. The most prominent peaks are labelled (in red).  In comparison, the background levels of commercial CaWO$_4$ crystals operated as detectors in CRESST II Phase 2 are shown. The  dashed red line indicates the rate of the crystal ``Daisy'' which has the lowest background rate among all commercial crystals ($\sim6$\,counts/[kg keV day]). The dashed black line shows the background of a typical commercial crystal, called ``VK31'' ($\sim30$\,counts/[kg keV day]). In that, the characteristic beta spectra of $^{227}$Ac and $^{210}$Pb clearly dominate.}
\label{fig:beta_spectrum}       
\end{figure}

The dominant part of events observed with CRESST-II detector modules are beta or gamma induced electron recoils which result in a highly populated beta/gamma band. At lowest energies ($E\lesssim10\,$keV), due to the finite resolution of the light detector there is a strong overlap with the ROI for dark matter search. Therefore, the beta/gamma background level is crucial for the sensitivity of CRESST-II detectors.\\
The low-energy spectra ($E\leq100$\,keV) of commercially available CaWO$_4$ crystals are usually dominated by two  intrinsic beta/gamma  background components \cite{Lang:electron_background}:
\begin{itemize}
\item Beta-decays of $^{210}$Pb which is part of the $^{238}$U chain. It decays into $^{210}$Bi with a half-life of 22.3\,years \cite{richard1999table}. In 84\% to all cases, an excited energy state of $^{210}$Bi with an energy of 46.5\,keV is populated. If the contamination is intrinsic to the crystal, both, the energy of the de-excitation gamma and that of the corresponding electron is detected in the crystal. These energy depositions cannot be resolved as separate events with cryogenic detectors and, hence, the two signals add. In many crystals, prominent $^{210}$Pb beta spectra are visible (decreasing in intensity from 46.5\,keV towards the Q-value of 63.5\,keV) which are evidence of  strong internal contaminations. If a $^{210}$Pb decay takes place on surfaces of materials outside the detector modules, only the de-excitation gamma is detected which results in a distinctive  peak at 46.5\,keV (external contamination).
\item Beta-decays of $^{227}$Ac which is part of the $^{235}$U   chain. It decays to $^{227}$Th  with a half-life of 21.8\,years \cite{richard1999table}. In 35\% (10\%) of the decays, an excited state of the daughter nucleus at the  energy level of 24.5\,keV (9.1\,keV) is populated which relaxes by gamma-emission to the ground state. Therefore, 3 beta-spectra are visible, each of them extending up to the Q-value of 44.8\,keV with two characteristic edges at 24.5\,keV and 9.1\,keV \cite{Lang:2009uh}.
\end{itemize}
In Fig. \ref{fig:beta_spectrum}, the characteristic features of $^{210}$Pb and $^{227}$Ac are clearly visible in the spectrum (black dashed line) of the  crystal ``VK31'' which is operated in CRESST-II Phase 2. The mean background rate in the ROI is about 30 counts/[kg\,keV\,day] which is a typical value for CaWO$_4$ crystals of this  supplier\footnote{The crystals were supplied by the General Physics Institute of the Russian Academy of Sciences (Moscow, Russia)}. The best such commercial crystal in terms of intrinsic radiopurity, called ``Daisy'', has an average rate of $\sim6$ counts/[kg keV day] (red dashed line in Fig. \ref{fig:beta_spectrum}).\\

For the first time, a detailed beta/gamma-background study of a TUM-grown crystal (TUM40) was performed in this paper. The bulk contamination and the average background rate could significantly be  reduced to 3.51/[kg\,keV\,day] in the ROI which is a reduction by a factor of 2-10 compared to commercial CaWO$_4$ crystals. The histogram in Fig. \ref{fig:beta_spectrum} shows the low-energy spectrum  of TUM40  from the first 29\,kg-days exposure of  CRESST-II Phase 2. Distinct gamma (X-ray) peaks appear above a rather flat  background which was not observed in commercial crystals. The dominant ones could be identified to originate from cosmogenic activation of W isotopes:
\begin{itemize}
\item Proton capture  on $^{182}$W (and a successive decay) can result in $^{179}$Ta which decays via EC to $^{179}$Hf with a half-life of 665\,d. The EC signature in our detectors is exactly the binding energy of the shell electrons of  $^{179}$Hf. In addition to a peak at 65.35\,keV (K-shell) which was reported earlier \cite{Lang:electron_background} also distinct peaks at 11.27\,keV (L1-shell), 10.74\,keV (L2-shell) and 2.60\,keV (M1-shell) could be identified here. The energies of all identified lines agree within errors with literature values  and are listed in Table \ref{tab:cosmo} with their corresponding activities. For the latter, the detection efficiency of TUM40 is considered (see \cite{Angloher:2014myn}). The K-shell EC line  of $^{179}$Ta has the highest rate:  $A_K=(277.2\pm15.7)\,\mu$Bq/kg. 
\item Proton capture on $^{183}$W can result in $^{181}$W which decays via electron capture (EC) to $^{181}$Ta with a half-life of 121\,d. We confirm the presence of a line at 74.02\,keV (K-shell + 6.2\,keV gamma) \cite{Lang:electron_background} with an activity of $A_K=(58.1\pm14.5)\,\mu$Bq/kg. To observe higher-order shell EC processes, present statistics is not yet sufficient\footnote{Since  CRESST-II Phase 2 is planned to run for $\sim2$\,years a further confirmation  of these peaks  is in reach by comparing the evolution of the event rate over time with the expected half-lifes.}.  
\end{itemize}  
In addition, gamma lines from external radiation can be identified: lines at 46.54\,keV  from $^{210}$Pb and, at higher energies, from $^{234}$Th, $^{226}$Ra, $^{212}$Pb, $^{208}$Tl, $^{214}$Bi, $^{228}$Ac and $^{40}$K. A peak from copper X-ray fluorescence at 8.05\,keV (K$_\alpha$) is observed since most of the material surrounding the detector is copper. The corresponding activities are listed in Table \ref{tab:cosmo}. The  contribution  of these lines to the background in the ROI is investigated in section \ref{sec:decomposition}. \\ 
There are no obvious features of beta spectra visible in the data of TUM40. However, the background contribution of beta/gamma decays from natural decay chains could be derived by a detailed analysis of the alpha spectra (see section \ref{sec:natural}) in combination with a dedicated MC simulation (see section \ref{sec:decomposition}). 
\begin{table}
\centering
\caption{Activities $A$ of the identified gamma-lines measured with the crystal TUM40 in 29\,kg-days of exposure during CRESST-II Phase 2.  External  gamma lines (ext.) as well as peaks from cosmogenically-induced EC decays  are listed (the shell of the captured electron is shown in brackets). }  
\label{tab:cosmo}       
\begin{tabular}{lcc}
\hline\noalign{\smallskip}
source & $E_{lit}$\,[keV] \cite{richard1999table}  &  $A$\,[$\mu$Bq/kg] \\
\noalign{\smallskip}\hline\noalign{\smallskip}
$^{179}$Ta (M1)& 2.6009 & 70.3$\pm$15.8\\
Cu X-ray (ext.)&8.048&27.1$\pm$14.3\\
$^{179}$Ta (L2)& 10.74&24.8$\pm$14.5\\
$^{179}$Ta (L1)& 11.271&202.2$\pm$16.0\\
$^{210}$Pb (ext.)&46.54&78.6$\pm$14.8\\
$^{179}$Ta (K)& 65.35&277.2$\pm$15.7\\
$^{181}$W (K)& 74.02&58.1$\pm$14.5\\
$^{234}$Th (ext.)&92.4&123.5$\pm$11.7\\
$^{226}$Ra (ext.)& 186.2&109.0$\pm$11.2\\
$^{212}$Pb (ext.)& 238.6&185.6$\pm$12.9\\
$^{208}$Tl (ext.)& 583.2&93.4$\pm$8.4\\
$^{214}$Bi (ext.)& 609.2&69.3$\pm$24.6\\
$^{228}$Ac (ext.)& 911.2&54.8$\pm$6.8\\
$^{228}$Ac (ext.)& 969.0&$36.3\pm$6.1\\
$^{40}$K (ext.)& 1432.7&41.4$\pm$6.2\\
$^{208}$Tl (ext.)& 2614.0&28.5$\pm$3.9\\
\noalign{\smallskip}\hline
\end{tabular}
\end{table}

\subsection{Surface-alpha background}\label{sec:surface}
\begin{figure}
\centering
\includegraphics[width=0.9\textwidth]{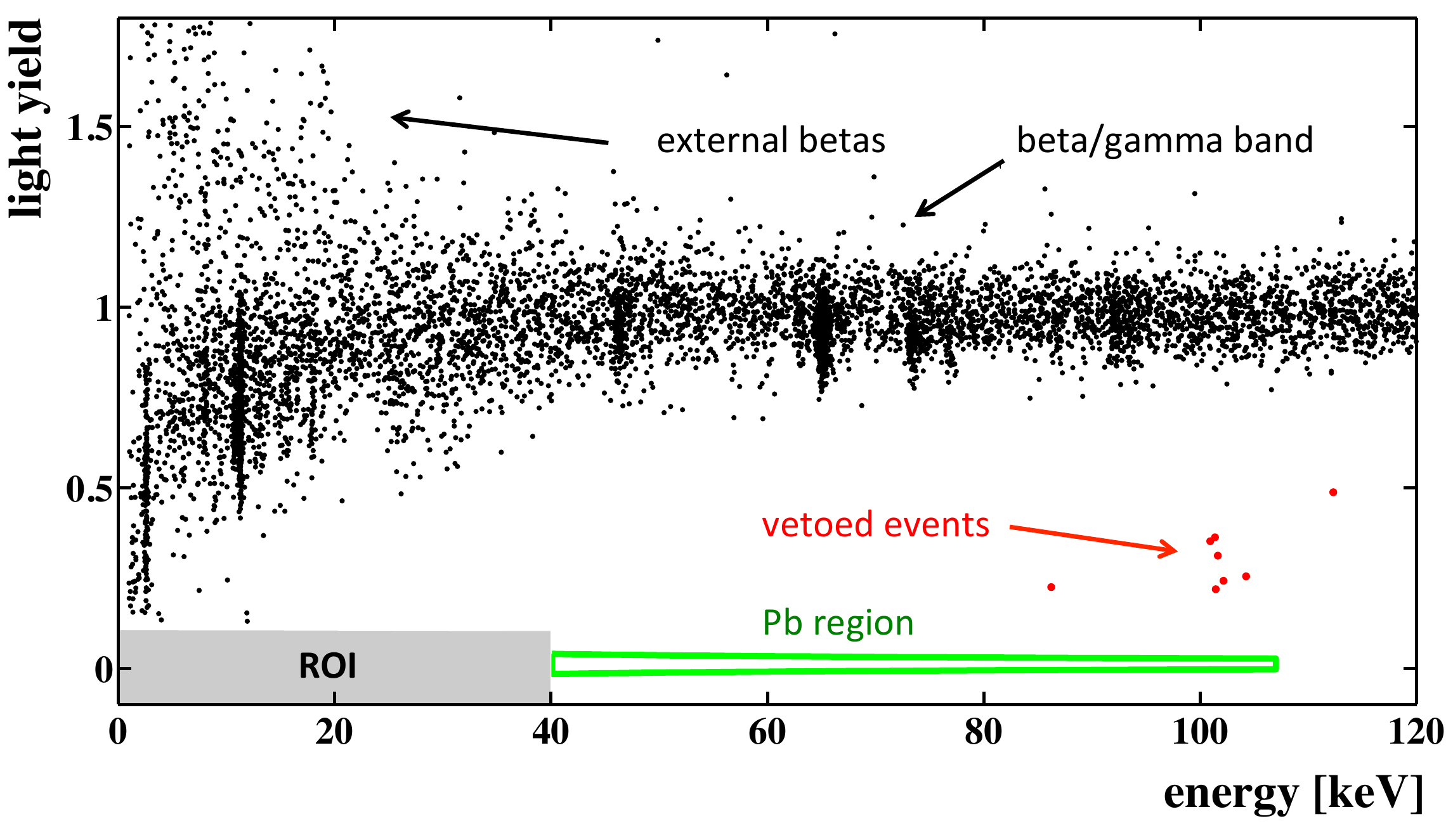}
\caption{Light yield vs. energy plot of the events (dots) recorded with TUM40 in CRESST-II Phase 2. At light yields of $\sim1$ the dominant band of beta/gamma events shows up. Due to light quenching \cite{birks1964theory} nuclear recoils have a reduced light yield. The ROI for dark matter search is shown in grey. Surface-alpha related $^{206}$Pb recoil background would be expected in the region enclosed by the green line. Due to the scintillation veto of the novel detector module, this area is free of events. All $^{206}$Pb recoils observed are shifted to higher light yields (red dots). External betas passing through the scintillating foil show up at higher light yields compared to events in the beta/gamma band. }
\label{fig:ly_plane}       
\end{figure}

A more detailed description of the crucial backgrounds related to surface alpha-events is given in \cite{strauss:2014part1}. Here, the basic results of that analysis are presented.\\
The data of CRESST-II detectors is often displayed in the light yield vs. energy plane, as shown in Fig. \ref{fig:ly_plane} for TUM40: At light yields of $\sim1$ the dominant beta/gamma band arises, almost horizontal at energies $\gtrsim50$\,keV and slightly decreasing (in light yield) towards smaller energies due to the scintillation properties of CaWO$_4$ \cite{Lang:2009uh}. The parametrisation of the event bands is given in \cite{strauss:2014part1}. The ROI for dark matter search, including all 3 nuclear recoil bands from threshold to 40\,keV\footnote{In \cite{Angloher:2014myn}, the 50\% acceptance bound for O recoils is defined as the upper light yield bound for the ROI. This convention is also used in this paper.}, is shown as a grey area in Fig. \ref{fig:ly_plane}. The reference region for $^{206}$Pb recoils induced by $^{210}$Po alpha decays is also indicated in the plot (area enclosed by the green line) at  light yields of $\sim0.01$ between 40\,keV and 107\,keV. In earlier runs of CRESST-II this region was populated: At the full kinetic energy of the Pb nucleus (decay on surfaces of surrounding material and of the crystal) a peak arose at $\sim103$\,keV with a exponential tail towards  lower energies. The latter comes from $^{210}$Po decays which are implanted in surrounding material. It was identified to originate from a contamination in the metal clamps holding the crystals (e.g. by Rn-implantation from air). The resulting spectrum of Pb-recoils was reproduced by a SRIM simulation \cite{Angloher:2012vn}.\\ 
In the first data of TUM40 acquired in CRESST-II Phase 2 no events are observed in the reference region for $^{206}$Pb recoils, while with the background level of Phase 1 \cite{Angloher:2012vn} $6.9\pm2.6$ events would be expected. Since in the novel detector module (see section \ref{sec:TUM40}) the crystal is completely surrounded by scintillating materials, the corresponding alpha with an energy of $\sim5.3$\,MeV  produces sufficient additional light to shift $^{206}$Pb recoils out of the nuclear recoil region. A population of such  vetoed  events is visible in Fig. \ref{fig:ly_plane} at light yields of 0.2-0.5 (red dots). This first data show the high efficiency of the surface-alpha event rejection realized with the new detector concept. No such backgrounds are expected within the sensitivity goals of CRESST-II Phase 2 with the final exposure \cite{strauss:2014part1}. \\
Furthermore, an additional background source most probably originating from contaminations on surfaces of materials outside the detector housing was identified in this paper. Electrons (e.g. from beta-emitters from natural decay chains) pass through the polymeric foil (thickness 50\,$\mu$m), lose some of their energy therein and produce additional light before getting absorbed by the CaWO$_4$ crystal. Dedicated measurements show that electrons, independent of their energy, produce a roughly fixed amount of scintillation light in the foil with an energy of $\mathcal{O}$(1\,keV). This explains the additional population of events above the beta/gamma band (labelled by ``external betas'' in Fig. \ref{fig:ly_plane}). Since an approximately fixed amount of  addition scintillation light  adds to the light produced by the beta in the crystal, the excess in light yield gets bigger towards lower energies. The analysis performed in section \ref{sec:decomposition} finds a significant contribution of this event population to the overall background in the ROI.

\subsection{Alpha background}\label{sec:natural}
\begin{figure}
\centering
\includegraphics[width=0.9\textwidth]{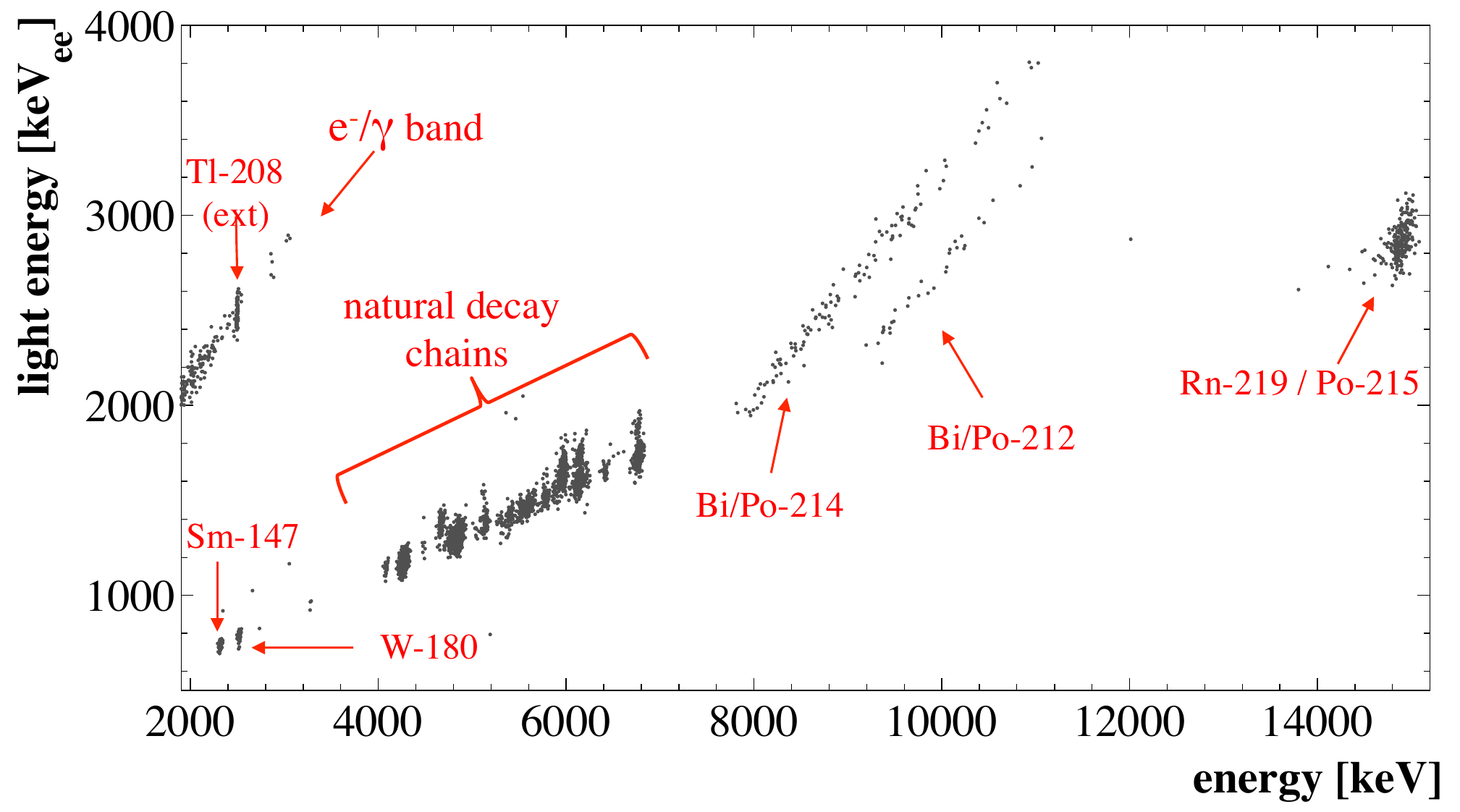}
\caption{Energy measured in the phonon detector plotted against the energy detected in the light detector in units of electron-equivalent energy (keV$_{ee}$). The  alpha lines are well separated from the beta/gamma band due to light quenching ($QF_\alpha\sim0.22$). Most of the discrete alpha lines of the natural decay chains have energies between 4 and 7\,MeV (see Fig. \ref{fig:histo_alpha}) except the $\alpha{-}\alpha$ ($^{219}$Rn/$^{215}$Po) and $\alpha{-}\beta$ ($^{214}$Bi/$^{214}$Po, $^{212}$Bi/$^{212}$Po) coincidences at higher energies. Lines from intrinsic  $^{180}$W and a contamination of $^{147}$Sm are also visible.}
\label{fig:alpha_spectrum_2d}       
\end{figure}

\begin{figure}
\centering
\includegraphics[width=0.9\textwidth]{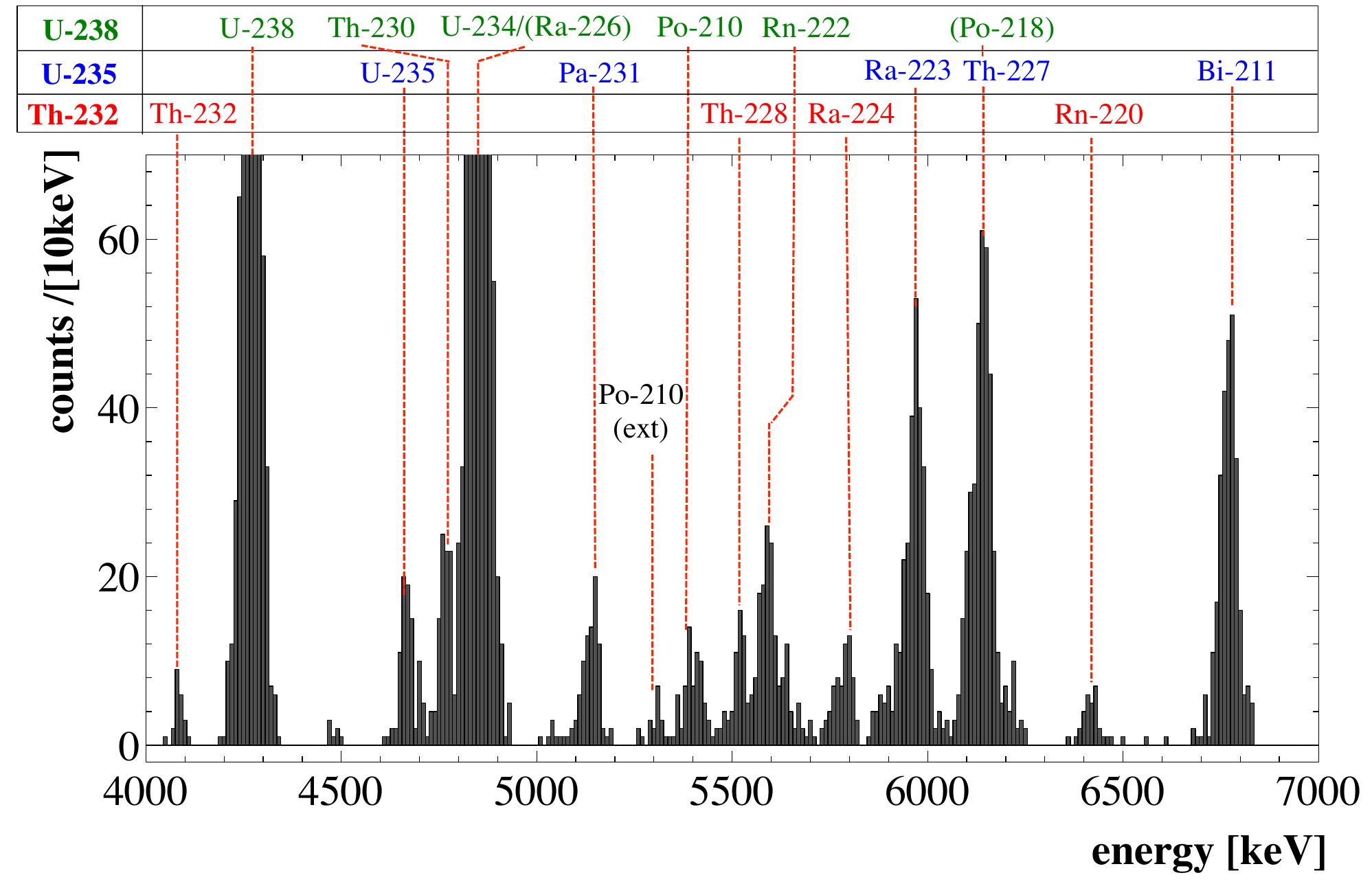}
\caption{Discrete alpha lines from the natural decay chains observed with the crystal TUM40 in an exposure of 29\,kg-days between 4 and 7\,MeV. The lines identified in the $^{238}$U,  $^{235}$U  and  $^{232}$Th chains are listed. In addition, an external $^{210}$Po line is visible where only the alpha energy (5.30\,MeV) and not the recoil of $^{206}$Pb (103\,keV) is detected. The individual activities are listed in Table \ref{tab:activities}. The peaks of  $^{238}$U and $^{234}$U completely dominate the spectrum (upper cut in histogram at 70 counts/[10\,keV]). }   
\label{fig:histo_alpha}    
\end{figure}

Due to the phonon-light technique of CRESST-II detectors, alpha events can be perfectly discriminated from other backgrounds, particularly, from the dominant beta/gamma populations. In case of an internal contamination, the full energy (Q-value) of a decay is measured. Therefore, the most obvious approach to  investigate intrinsic radioactivity of  CaWO$_4$ crystals is the analysis of discrete alpha lines  which arise at energies $\gtrsim2\,$MeV far off the ROI.
Distinct alpha peaks from intrinsic contamination of the natural decay chains as well as of radioactive rare-earth metals (e.g. $^{147}$Sm) and the isotope $^{180}$W are observed.\\ 
In TUM40 decays from the three natural decay chains $^{238}$U,  $^{235}$U  and  $^{232}$Th are observed.  All individual decays  are listed with their respective half-life and Q-values ($E_{lit}$) in Table \ref{tab:activities}. 
\begin{table}
\caption{Radioactive decays (alpha and beta) of the natural decay chains ($^{238}$U,  $^{235}$U  and  $^{232}$Th) from intrinsic contamination observed with the crystal TUM40. The first 29\,kg-days of exposure of CRESST-II Phase 2 were used for the analysis. The Q-values ($E_{lit}$) from literature \cite{richard1999table} of  the individual reactions   are compared with the energies ($E_{obs}$) observed in the detector.  The corresponding activities $A$ are given with $1\sigma$ errors. The branching ratios ($br$) of the individual reactions are listed. The symbol (a) indicates beta decays which cannot  be identified individually but are in equilibrium with short-lived alpha decays. Alpha-lines marked with (b) have a significant overlap with stronger alpha sources. The symbol (c) indicates fast decays which cannot (or only partly) be distinguished from the proceeding decay within detector resolution. The total activity of the identified intrinsic alpha sources is $A_{tot,\alpha}=3.08\pm$0.04\,mBq/kg.  }
\label{tab:activities}       
\centering
\begin{tabular}{llllllll}
\hline\noalign{\smallskip}
chain & parent & mode & $br$\,[\%] &   half-life & $E_{lit}$\,[keV] & $E_{obs}$\,[keV] &  $A$\,[$\mu$Bq/kg] \\
\noalign{\smallskip}\hline\noalign{\smallskip}
$^{238}$U &$^{238}$U & $\alpha$ & 100   &$4.47\cdot10^9$\,y & 4270 & 4271       &($1.01{\pm}0.02){\cdot}10^{3}$ \\
          &$^{234}$Th& $\beta^-$& 100   &24.1\,d 			   & 273 & 0-273 &(a)         \\
          &$^{234}$Pa$^\ast$& $\beta^-$& 100   &1.17\,min	       & 2197&  0-2197 &   (a)     \\
          &$^{234}$U & $\alpha$ & 100   &$2.45\cdot10^5$\,y & 4858 & 4853       &($1.08{\pm}0.03){\cdot}10^{3}$ \\
          &$^{230}$Th& $\alpha$ & 100   &75.4\,y            & 4770 & 4771       &$55.8\pm$5.4                   \\
          &$^{226}$Ra& $\alpha$ & 100   &1.60\,y            & 4871 & 4853\,(b)    &$43.0\pm$9.9                  \\
          &$^{222}$Rn& $\alpha$ & 100   &3.82\,d            & 5590 & 5592       &$38.1\pm$4.9                  \\
          &$^{218}$Po& $\alpha$ & 99.98 &3.10\,min          & 6115 & 6139       &$43.1\pm$9.9            \\
                    &$^{214}$Pb& $\beta^-$ & 100 &26.8\,min          & 1023 & 0-1023       & (a)          \\
          &$^{214}$Bi& $\beta^-$& 99.98 &19.9\,min          & 3272 &\multirow{2}{*}{7800-11000}&\multirow{2}{*}{$47.4\pm$4.9}   \\
          &$^{214}$Po& $\alpha$ & 100   &0.164\,ms\,(c)     & 7883 &            &                                \\
          &$^{210}$Pb& $\beta^-$& 100   &22.3\,y            & 63.5 & 0-63.5     &$7^{+36}_{-7}$                 \\
		  &$^{210}$Bi& $\beta^-$& 100   &5.01\,d            & 1163 & 0-1163     &(a)                  \\          
          &$^{210}$Po& $\alpha$ & 100   &138\,d             & 5407 & 5403       &$17.8\pm$4.0                   \vspace{0.2cm} \\
$^{235}$U &$^{235}$U & $\alpha$ & 100   &$7.04\cdot10^8$\,y & 4678 & 4671       &$39.5\pm$4.4                   \\
 	&$^{231}$Th & $\beta^-$ & 100   &25.52\,h & 389.5 & 0-389.5       &(a)                   \\
          &$^{231}$Pa& $\alpha$ & 100   &$3.27\cdot10^4$\,y & 5150 & 5139       &$23.2\pm$4.4                 \\
          &$^{227}$Ac& $\beta^-$& 98.62 &21.8\,y          & 44.8 & 0-44.8     &$98\pm$20                   \\
          &$^{227}$Th& $\alpha$ & 100   &18.7\,d          &   6147 & 6139     &$105\pm$19                    \\
          &$^{223}$Ra& $\alpha$ & 100   &11.4\,d          &   5979 & 5968     &$104\pm$7                      \\
          &$^{219}$Rn& $\alpha$ & 100   &3.96\,s          & 6946   & \multirow{2}{*}{14900}             &\multirow{2}{*}{$107\pm$7}  \\
          &$^{215}$Po& $\alpha$ & 100   &1.78\,ms\,(c)         &  7527  &         &                               \\
      	  &$^{211}$Pb& $\beta^-$& 100   &36.1\,min          &  1373 &  0-1373     &(a)                     \\         
          &$^{211}$Bi& $\alpha$ & 100   &0.51\,s          &  6751 & 6771      &$105\pm$7                      \\
          &$^{207}$Tl& $\beta^-$ & 100   &4.77\,min         &  1423  & 0-1423      & (a)         \vspace{0.2cm}\\
$^{232}$Th&$^{232}$Th& $\alpha$ & 100   &$1.40\cdot10^{10}$\,y& 4083& 4084      &$9.2\pm$2.3                  \\
		  &$^{228}$Ra& $\beta^-$ & 100   &5.75\,y          &  45.9 & 6.7-45.9      &(a)                 \\
		  &$^{228}$Ac& $\beta^-$ & 100   &6.15\,h          &  2127& 58-2127      &(a)                 \\
		  &$^{228}$Th& $\alpha$ & 100   &1.91\,y          &  5520 & 5518      &$15.2\pm$4.1              \\
		  &$^{224}$Ra& $\alpha$ & 100   &3.63\,d          &  5789 & 5788      &$19.8\pm$8.1                  \\
		  &$^{220}$Rn& $\alpha$ & 100   &55.6\,s          &  6404 & 6414      &$8.4\pm$3.4                     \\
		  &$^{216}$Po& $\alpha$ & 100   &0.145\,s         &  6906 & -  & 0                              \\
		  &$^{212}$Pb& $\beta^-$ & 100   &10.64\,h          &  573.7& 0-573.7      &(a)                 \\
		  &\multirow{2}{*}{$^{212}$Bi}& $\alpha$& 35.94 &\multirow{2}{*}{60.6\,min}        &  6208 & 6216       &$7.7^{+8.9}_{-7.7}$     \\
		  && $\beta^-$& 64.06 &        &  2252 &\multirow{2}{*}{8900-11200}       &\multirow{2}{*}{$15.8\pm$2.8}  \\
          &$^{212}$Po& $\alpha$ & 100   &299\,ns\,(c)         & 8955 &                &                         \\
          &$^{208}$Tl& $\beta^-$ & 100   &3.01\,min          &  5001& 3197-5001      &(a)                  \\
\noalign{\smallskip}\hline
\end{tabular}
\end{table}
The detailed study of the intrinsic alpha contaminations performed in this work enables to derive all relevant beta-decay rates. This provides a quantification of the intrinsic contamination at low-energies for the first time.\\
The large dynamic range and the discrimination capability of CRESST-II detectors for alpha events is illustrated in Fig. \ref{fig:alpha_spectrum_2d}. The  energy measured in the phonon detector is plotted versus the (electron-equivalent) energy measured in the light detector with the module TUM40 in 29\,kg-days of exposure. Due to light quenching the alpha band is well-separated from the  beta/gamma band. Besides the lines of $^{147}$Sm and $^{180}$W, all peaks can be attributed to alpha decays from natural decay chains clustering at energies between 4 and 7\,MeV. The total alpha activity of  intrinsic contamination is $A_{tot}=(3.08\pm$0.04)\,mBq/kg to which $^{238}$U ($(1.01\pm$0.02)\,mBq/kg) and  $^{234}$U  ($(1.08\pm$0.03)\,mBq/kg) from the $^{238}$U chain  are the dominant contribution. The histogram in Fig. \ref{fig:histo_alpha} shows the identified alpha peaks of lower intensities ($(\sim10-100)\,\mu$Bq/kg) between 4 and 7\,MeV.  At even higher energies beta-alpha coincident events are visible in Fig. \ref{fig:alpha_spectrum_2d} ($^{214}$Bi/$^{214}$Po and $^{212}$Bi/$^{212}$Po at $\sim7.8-12\,$MeV). These subsequent decays are too fast to be disentangled as separate events, hence both signals sum up and characteristic continuous bands arise (see \cite{muenster_2014} for details). Similarly, the alpha-alpha coincidence $^{219}$Rn/$^{215}$Po shows up at the combined energy of $\sim14.9\,$MeV. The measured $^{180}$W activity of  $(36\pm9$)$\,\mu$Bq/kg agrees well with the  rate of $(31\pm8$)$\,\mu$Bq/kg  obtained by a previous measurement \cite{Cozzini:2004vd} which cross-checks the analysis performed.\\
All alpha particles originating from natural decay chains are observed in this data and  are listed in Table \ref{tab:activities} with their individual activities. All alpha decays, short-lived with respect to the measuring time ($\sim0.5$\,y) agree in rate as expected. The equilibrium of the  $^{238}$U chain is broken at the long-lived isotopes $^{234}$U  \footnote{A broken equilibrium at $^{210}$Pb and $^{210}$Po ($^{238}$U chain) is possible, however, can not be proven here due to limited statistics.}  and that of the   $^{235}$U  chain at $^{231}$Pa. 
This can originate, e.g., from chemical separation processes during the production steps. All measured intensities are consistent with the individual decay chains (a reduced detection efficiency for the decays $^{220}$Rn and $^{216}$Po due to pile-up effects has to be considered, for details see \cite{muenster_2014}).        

\section{Monte Carlo based decomposition of the background} \label{sec:decomposition}
\begin{figure}
\centering
\includegraphics[width=\textwidth]{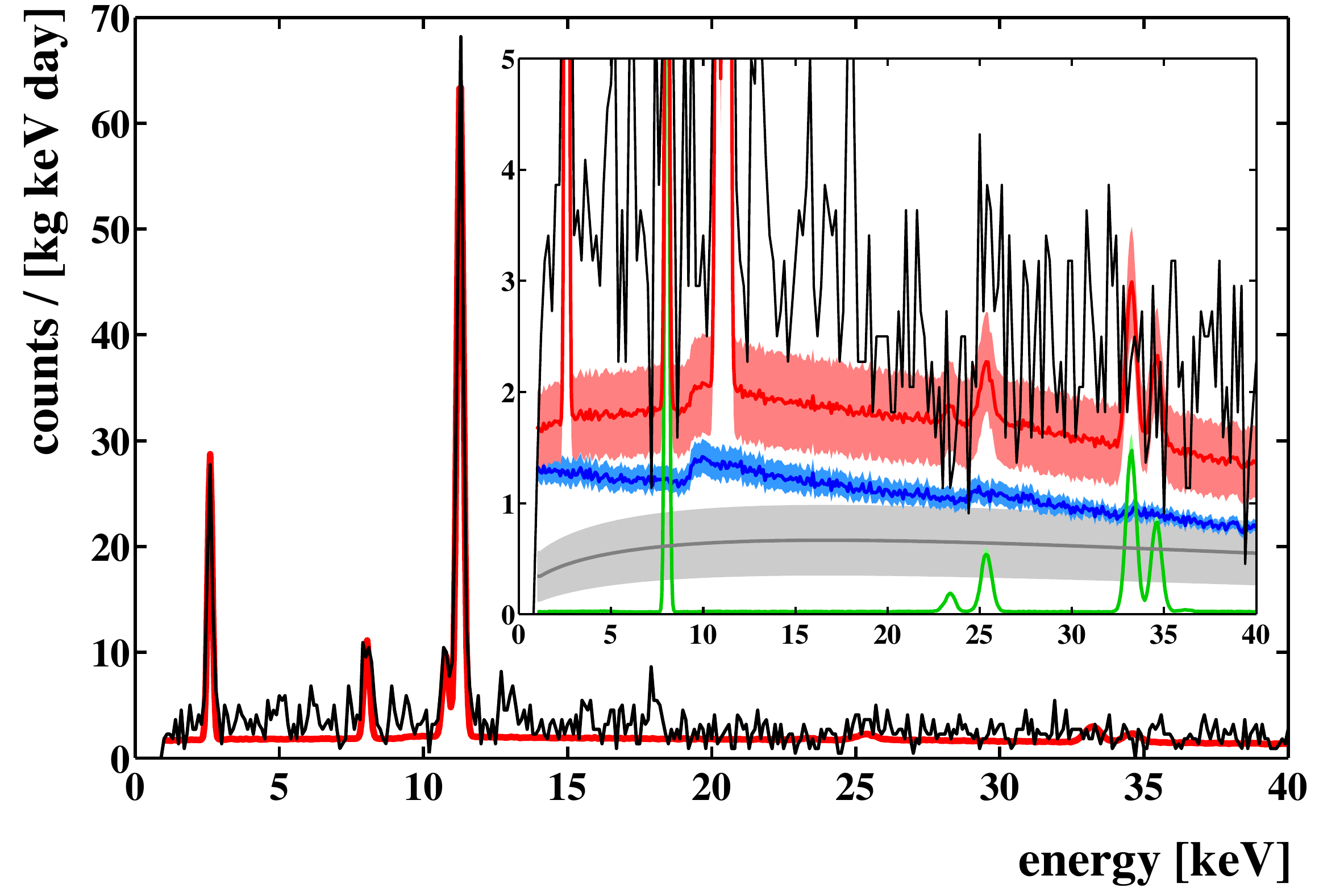}
\caption{Histogram of the events in the ROI (black line) recorded with TUM40 in CRESST-II Phase 2. The red line indicates the sum of all identified background sources with the dominant peaks from cosmogenic activation (2.6\,keV, 10.7\,keV, 11.3\,keV) and the Cu X-ray line (8.0\,keV). Inset: Decomposition of the  background based on MC simulation (see text). The contributions of external gamma radiation (green), external betas (grey) and intrinsic beta/gamma radiation from natural decay chains (blue) are shown. The sum of these components (plus gamma peaks) are shown in red. The individual 1-$\sigma$ error bands are depicted in the corresponding colour. The identified backgrounds explain  $\sim$70\% of the observed events.   }
\label{fig:beta_spectrum_fitted}    
\end{figure}

\begin{figure}
\centering
\includegraphics[width=0.9\textwidth]{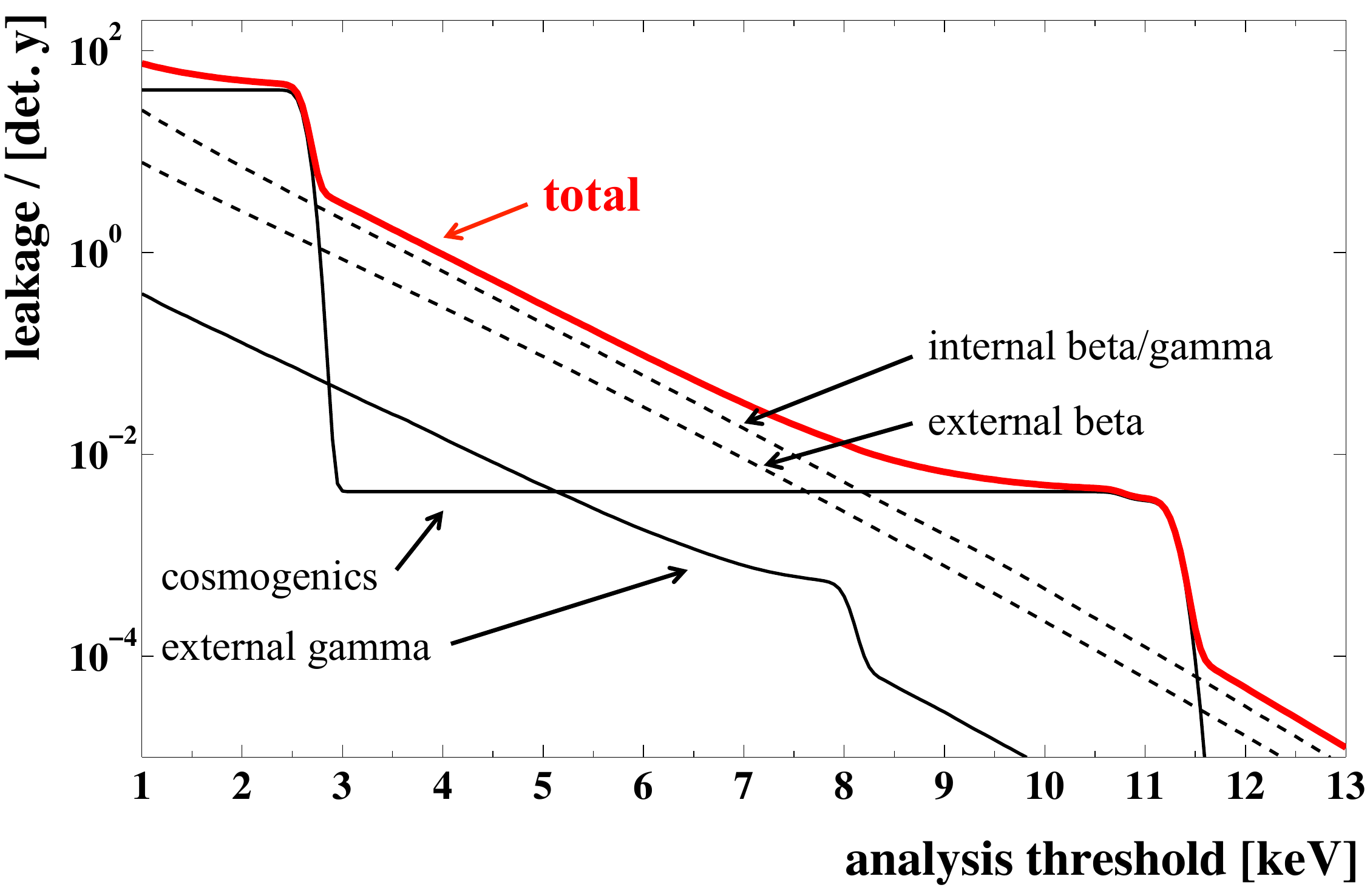}
\caption{Expected cumulative leakage from the beta/gamma band into the ROI (1 to 40keV) plotted against the analysis threshold. The calculation is based on the background level of the detector TUM40 in CRESST-II Phase 2. The decomposition of the background into individual components is explained in detail in section \ref{sec:decomposition}. } 
\label{fig:leakage}   
\end{figure}

By the detailed alpha analysis presented in section \ref{sec:natural} the intensities of all beta decays from the natural decay chains $^{238}$U,  $^{235}$U  and  $^{232}$Th can be derived since they are in equilibrium with at least one alpha decay (see Table \ref{tab:activities}). These measured activities  were used as input for a dedicated GEANT4 MC simulation (version 10.1) \cite{Agostinelli2003250,1610988}  which allows to derive the beta/gamma spectra in the ROI for dark matter search. To model the production, propagation, and absorption of beta/gamma particles in the detector down to 250eV, the Livermore low-energy electromagnetic models were used. The electromagnetic model implements  photo-electric effect, Compton scattering, Rayleigh scattering, conversion in e$^+$e$^-$ pairs, ionisation, and bremsstrahlung production.\\  
The results of this data-based simulation is shown in Fig. \ref{fig:beta_spectrum_fitted} (inset): The blue curve shows the sum of all beta/gamma events from natural decay chains. The 1-$\sigma$ error band (light blue) is a combination of the statistical error of the simulation and the uncertainty of the experimentally determined activities of the beta emitters. This contribution has  an activity of $A_{1-40}=494.2\pm48.4\,\mu$Bq/kg in the ROI which corresponds to a mean rate of $3.51\pm0.09$ counts/[kg\,keV\,day]. For the first time, the contribution of the intrinsic beta/gamma emitter could be disentangled. It corresponds to $(30.4\pm2.9)$\% of the total events observed. The main contributions originate from $^{234}$Th (346\,$\mu$Bq/kg),  $^{227}$Ac (93\,$\mu$Bq/kg), $^{234}$Pa (35\,$\mu$Bq/kg) and $^{228}$Ra decays (9\,$\mu$Bq/kg). The characteristic edges at $\sim9$\,keV and $\sim24$\,keV originate from the contribution of the $^{227}$Ac spectrum (see section \ref{sec:low_energy_spectrum}).  The values of all relevant beta emitters are listed in Table \ref{tab:leakage}.\\        
Furthermore, the response of the detector to external gamma radiation is studied with a dedicated MC simulation. The intensity of the individual components is scaled such to match the observed external gamma peaks (see section \ref{sec:low_energy_spectrum}). All identified external gamma lines which are listed in Table \ref{tab:cosmo} are included in the study. The result is shown as in Fig. \ref{fig:beta_spectrum_fitted} (inset) as a green line with the corresponding 1-$\sigma$ error band (light green). The only peak in the ROI identified as to originate from external radiation is the Cu X-ray peak at 8.0\,keV.   The continuous Compton background from external sources (peaks at higher energies) is negligible in the ROI ($\lesssim0.03$ events/[kg keV day]), however, W-escape peaks originating from the $^{234}$Th line  at around 92.4\,keV contribute significantly (($62.2\pm7.8$)\,$\mu$Bq/kg).  X-rays from W (K$_{\alpha1}$: 59.3\,keV, K$_{\alpha2}$: 58.0\,keV, K$_{\beta1}$: 67.2\,keV and K$_{\beta2}$: 69.1\,keV \cite{richard1999table}) can escape the CaWO$_4$ crystal giving rise to 4 distinct escape peaks between 23 and 35\,keV. The statistics of the recorded data is not yet sufficient to resolve these peaks, but as shown below, the total observed spectrum is compatible with this prediction.\\
In addition to that, a background component from external betas has been identified within this work. As explained in section \ref{sec:surface}, electrons passing through the polymeric foil produce additional scintillation light before hitting the CaWO$_4$ crystal which shifts these events to higher light yields (see Fig. \ref{fig:ly_plane}). A phenomenological model to account for this event population was developed in \cite{schmaler_Phd}. Measurements show that these events can be nicely described  by an exponential distribution decreasing towards higher energies and higher light yields.  To obtain the observed event distribution, the spectrum has to be convolved with the resolution of the light measurement. A detailed description of the likelihood fit   can be found in the literature  \cite{strauss:2014part1,schmaler_Phd}. The result of the likelihood fit to the data of TUM40 is shown in Fig. \ref{fig:beta_spectrum_fitted} (inset): The grey curve accounts for the rate of external betas most probably originating from beta emitters on surfaces surrounding the detectors. The corresponding 1-$\sigma$ error bounds are depicted in light grey. An activity of $(274\pm137.6)\,\mu$Bq/kg which corresponds to $(16.9\pm8.4)$\% of the total rate is attributed to external betas.\\ 
The EC-peaks (at 2.60\,keV, 10.74\,keV and 11.27\,keV) originating from cosmogenic activation of W isotopes (see section \ref{sec:low_energy_spectrum}) account for $(290.7\pm17.0)$\,$\mu$Bq/kg ($(17.9\pm1.0)$\% of the total rate).\\
The red curve in Fig. \ref{fig:beta_spectrum_fitted} (main frame and inset) shows the combined spectrum of all identified backgrounds. The inset shows the corresponding 1-$\sigma$ error bounds (light red). A total activity of          ($1121.8\pm146.6$)\,$\mu$Bq/kg is found which explains ($69.0\pm9.0$)\% of the data. While at energies above $\sim20$\,keV the sum of identified sources almost completely reproduces the data, at lower energies a significant part of the spectrum remains unexplained. There are slight hints for additional gamma peaks (e.g at $\sim18$\,keV), for the identification of which, however, more statistics is necessary.\\  
\begin{table}
\centering
\caption{Activities ($A_{1-40}$) of the identified background sources in the $e^-/\gamma$-band between 1 and 40\,keV  and the expected leakage $L_1$ ($L_{12}$) per detector ($m\,{=}\,245$\,g) and year into the region-of-interest (ROI) assuming an analysis threshold of 1\,keV (12\,keV). The activities of the relevant $\beta^-$-decaying isotopes from natural decay chains  are derived by a MC-based analysis (see text). The sum of all intrinsic beta decays ($\Sigma$ internal $\beta/\gamma$), of the cosmogenic activation lines ($\Sigma$ cosmogenics), of the external $\gamma$ radiation ($\Sigma$ external $\gamma$ ) and of the external $\beta$ radiation ($\Sigma$ external $\beta$) are listed. The sum of all identified sources is compared to the activity observed with in the data of CRESST-II Phase 2. For   $L_1$ and $L_{12}$ the total errors are of $\mathcal{O}$(10\%). (a) indicates an extrapolation to the assumed exposure of 1 detector-year. }
\label{tab:leakage}       
\begin{tabular}{lccc}
\hline\noalign{\smallskip}
source & $A_{1-40}$\,[$\mu$Bq/kg] & $L_{1}$/[det.-y] &  $L_{12}$/[det.-y] \\
\noalign{\smallskip}\hline\noalign{\smallskip}
$^{234}$Th & $346.2\pm12.3$ & 18.1 & $2.0\cdot10^{-5}$\\
$^{234}$Pa$^\ast$ & $35.0\pm5.7$ & 1.6 & $2\cdot10^{-6}$\\
$^{214}$Pb & $(8.3\pm0.4)\cdot10^{-1}$ & $3.7\cdot10^{-2}$ & $<10^{-6}$\\
$^{210}$Pb & $1.0\pm0.1$ & $8.6\cdot10^{-2}$ & $<10^{-6}$\\
$^{210}$Bi & $(3.6\pm0.4)\cdot10^{-1}$ & $1.6\cdot10^{-2}$ & $<10^{-6}$\\
$^{231}$Th & $1.1\cdot10^{-2}$ & $2.5\cdot10^{-4}$ & $<10^{-6}$\\
$^{227}$Ac & $92.9\pm1.8$ & 5.6 & $9.0\cdot10^{-6}$\\
$^{211}$Pb & $4.2\pm0.1$ & $1.9\cdot10^{-1}$ & $<10^{-6}$\\
$^{207}$Tl & $3.9\pm0.6$ & $1.7\cdot10^{-1}$ & $<10^{-6}$\\
$^{228}$Ra & $8.9\pm0.2$ & $8.3\cdot10^{-4}$ & $1\cdot10^{-6}$\\
$^{212}$Pb & $(2.4\pm0.4)\cdot10^{-1}$ & $1.1\cdot10^{-2}$ & $<10^{-6}$\\
Cu X-ray (ext.) & $27.1\pm5.1$ & $4.4\cdot10^{-4}$ & -\\
\noalign{\smallskip}\hline\noalign{\smallskip}
$\Sigma$ internal $\beta/\gamma$  & $494.2\pm48.4$ & 25.7 & $3.2\cdot10^{-5}$\\
$\Sigma$ cosmogenics & $290.7\pm17.0$ & 40.8 & $<10^{-6}$\\
$\Sigma$ external $\gamma$ & $62.2\pm7.8$ & $3.9\cdot10^{-1}$ & $1\cdot10^{-6}$ \\
$\Sigma$ external $\beta$ & $274.7\pm137.6$ & 7.8 & $1.6\cdot10^{-5}$ \vspace{0.2cm} \\
$\Sigma$ all identified & $1121.8\pm146.6$ & 74.7 & $4.9\cdot10^{-5}$\\
\noalign{\smallskip}\hline\noalign{\smallskip}
total observed & $1625.1\pm40.3$ & 108.3 (a) & $7.1\cdot10^{-5}$ (a)\\
\noalign{\smallskip}\hline
\end{tabular}
\end{table}
To investigate the relevance of the individual background components for dark matter search, their leakage into the nuclear-recoil bands is calculated based on the  background studies developed in this work. The number of events  leaking into the ROI (to below the 50\% acceptance bound for O-recoils \cite{Angloher:2014myn}) is calculated for the beta/gamma spectrum from natural decay chains, external gammas, the cosmogenic activation lines and the contribution of external betas. In Table \ref{tab:leakage} the cumulative leakage of the individual components into the ROI is listed for an analysis threshold of 1\,keV ($L_1$) and 12\,keV ($L_{12}$). The cumulative leakage  is graphically illustrated in Fig. \ref{fig:leakage} as a function of the analysis threshold. For an analysis threshold of 1\,keV a leakage of $\mathcal{O}(10^2)$ events per detector and year is expected. Here the contaminations due to cosmogenic activation give the highest contribution. Above ${\sim}12\,$keV, the leakage drops by $\sim2$ orders of magnitude and is then dominated   by intrinsic beta/gamma events.  The calculation shows that an experiment with a threshold of 12\,keV would be background-free up to an exposure of ${\sim}3\,$ton-years.       

\section{Conclusions and Outlook}
TUM40 operated in the new detector housing  has reached unprecedented background levels. Using CaWO$_4$ sticks to hold the target crystal, a fully-scintillating inner detector housing is realized and backgrounds from surface-alpha decays are rejected with high efficiency.  A phonon trigger threshold of $\sim0.60$\,keV and a resolution of $\sigma{=}(0.090{\pm}0.010)\,$keV (at 2.60\,keV) are reached with TUM40. By using a CaWO$_4$ crystal produced at the TUM, the intrinsic background rate was reduced to the lowest level reported for CRESST CaWO$_4$ detectors: on average $3.51\pm0.09$ beta/gamma events per kg\,keV\,day in the ROI (1-40\,keV) and    a total alpha activity from natural decay chains of $A_{tot,\alpha}=3.08\pm$0.04\,mBq/kg. In this paper, a detailed alpha analysis was performed which allowed to derive the activities of all decaying isotopes  of the natural decay chains. Based on these results, a GEANT4 MC simulation was set up to investigate the contribution of intrinsic beta/gamma backgrounds in the ROI for dark matter search (1-40\,keV). An activity of $\sim494\,\mu$Bq/kg was found which corresponds to $\sim30$\% of the total event rate. The MC simulation also shows the contribution of events originating from external gamma radiation. An activity of $62.2\,\mu$Bq/kg ($\sim4$\% of total) is found in the ROI. Furthermore, the study shows that 17\% of the background most probably is due to  electrons originating from surfaces surrounding the detector and passing through the scintillating housing.  Distinct gamma lines are observed in the low-energy spectrum at energies $<80\,$keV,  originating from cosmogenic activation of W isotopes. The isotopes $^{182}$W and $^{183}$W,  and subsequent EC-reactions of $^{179}$Ta and $^{181}$W are identified as sources. The resulting  activity in the ROI is 290.7\,$\mu$Bq/kg ($\sim18$\% of the total event rate).\\
For the first time, the background in the ROI of a CRESST-II detector module could be decomposed in a comprehensive way.
The analysis performed suggests that the background of TUM40 is - unlike previously assumed - not exclusively limited  by intrinsic contaminations from natural decay chains, but cosmogenic activation and external sources play  an important role. The knowledge of the background origin is crucial for the future R\&D activities of CRESST concerning background reduction.  \\
The sensitivity of CRESST-II detectors for dark matter search was investigated by calculating the leakage of beta/gamma events into the ROI. For analysis thresholds $>12\,$keV the present background level results in a total expected leakage of $<10^{-4}$ events per detector and year. This suggests that even for a ton-scale experiment and typical measuring times of $\mathcal{O}(1$\,year) such detectors would not be limited by beta/gamma leakage and, if no additional background sources contribute, would be background-free for high WIMP-mass searches ($\gtrsim5$\,GeV/c$^2$). The goals of the future EURECA experiment \cite{Angloher201441}, namely  to reach the level of $10^{-10}$\,pb for the spin-independent WIMP-nucleon scattering cross-section (for WIMP masses of $\sim50$\,GeV/$c^2$), can be achieved by TUM40-type detectors using state-of-the-art CRESST technology.\\
Due to the low thresholds achieved, CRESST detector are  most suited for the detection of low-mass WIMPs ($\lesssim5$\,GeV/c$^2$). The potential has been recently demonstrated by the first data of TUM40 in CRESST-II Phase 2 which sets stringent limits on the WIMP-nucleon scattering cross-section and explores new parameter space below WIMP masses of 3\,GeV/c$^2$ \cite{Angloher:2014myn}. The study performed in this paper shows that for an analysis threshold of 1\,keV a total leakage of $\sim10^2$ events is expected per detector and year. This agrees nicely with the number of events observed in the ROI during the first 29\,kg-days of data acquired with TUM40 and suggests that no other background sources contribute (as e.g. neutrons). To achieve the future sensitivity goals for low-mass WIMP search \cite{cresst3_strategy}, background levels must be further reduced. External radiation can be diminished by cleaning of material which are in the vicinity of the detectors (e.g. re-etching of copper surfaces) and additional shielding. To reduce the crystal-intrinsic backgrounds, the  crystal production at TUM is key. We plan to significantly reduce  the cosmogenic activation of W isotopes  by using only screened and assayed materials, and to shorten the exposition to cosmic rays during transport and storage. It was found, that the growth procedure itself segregates certain radioactive isotopes significantly. Comparing the contamination of $^{226}$Ra in TUM40 ($(43.0{\pm}9.9)\,\mu$Bq/kg, see Table \ref{tab:activities}) to that in the raw materials ($\mathcal{O}$(10\,mBq/kg) \cite{muenster_2014}), a purification  by almost 3 orders of magnitude is achieved. Therefore, a multiple re-crystallization of CaWO$_4$ crystals is a promising technique to decrease both, cosmogenic activation  and contaminations from natural decay chains.  


\acknowledgments

This research was supported by the DFG cluster of excellence: Origin and Structure of the Universe, the DFG Transregio 27: Neutrinos and Beyond, the Helmholtz Alliance for Astroparticle Phyiscs, the Maier-Leibnitz-Laboratorium (Garching), the STFC (United Kingdom) and by the BMBF: Project 05A11WOC EURECA-XENON. We are grateful to LNGS for their generous support of CRESST, in particular to Marco Guetti for his constant assistance.

\bibliographystyle{JHEP}       
\bibliography{quenching_bibtex}

\end{document}